# Theoretical investigation of structural, electronic properties and half metallic ferromagnetism in $Ca_{1-x}Ti_xS$ ternary alloys

Meryem Ziati [1, *] and Hamid Ez − Zahraouy [1]

[1] Laboratory of Condensed Matter and Interdisciplinary Sciences department of physics, Faculty of Sciences, Mohammed V University, Rabat – Morocco.

*Corresponding author: meryem.ziati.um5@gmail.com

**Abstract**

In this research paper, we investigated the structural, electronic, and magnetic features of titanium atom substituting calcium atom in rock-salt structure of CaS to explore the new dilute magnetic semiconductor compounds $Ca_{1-x}Ti_xS$. The calculations are carried out using the full potential-linearized augmented plane wave (FP-LAPW) method based on spin-polarized density functional theory (SP-DFT), implemented in WIEN2k code. The exchange and correlation potential are described by the generalized gradient approximation (PBE-GGA) and Tran-Balaha modified Becke-Johnson exchange potential (TB − mBJ). The stability of $Ca_{1-x}Ti_xS$ ternary alloys in ferromagnetic state is provided by the total energy released from the optimized structures and defect formation energies. The classical model of Heisenberg is employed to estimate Curie temperature of these compounds. It is found that the room temperature ferromagnetism achieved at low concentrations. The studied materials exhibit half-metallic ferromagnetic demeanor. The half metallic gaps ($G_{HM}$) are the extremely significant factors to consider for spintronic applications. The insertion of impurity significantly decreased the value of $G_{HM}$ due the broadening of Ti − 3d states in the gap of the minority spin. Furthermore, to evaluate the effects of the exchange splitting process, the p − d exchange splitting $\Delta E_C$, $\Delta E_v$ and the exchange constants $N_{0\alpha}$, $N_{0\beta}$ are predicted.

## 1- Introduction

Semiconductors are materials which have a conductivity between conductors (metals) and non-conductors or insulators. This type of materials can be pure elements, such as silicon (Si) or germanium (Ge), or compounds such as gallium arsenide (GaAs) [1] or cadmium selenide (CdSe) [2]. To engineer their band structures and expand their usefulness, small amounts of impurities are added to pure semiconductors causing a large change in their electronic performance. This process called 'doping' or 'substituting'. In general terms, transition metal (TM) doping such us (V, Cr, Mn, Fe, Co and Ni) [3] usually generates the new bands inside the semiconductor band gap and induces magnetic properties. This category of high spin polarized materials is called Diluted Magnetic Semiconductors (DMSs) [4].

DMSs materials have garnered enormous interest in theoretical and experimental studies due to their attractive physical properties such us magneto-electronics and magneto-transport [5]. They are mostly referred to spin transport electron technology due to the high Curie temperatures (Tc) [6] and usually exhibit the half metallic ferromagnetic (HMF) demeanor [7]. Half metals act as a conductor to electrons of one spin orientation, but as a semiconductor or insulator to those of the opposite orientation with 100% spin-polarized at Fermi level [8-9]. Hence, DMS could be regarded as a potential candidate material for various technological applications, including the manufacture of photovoltaic devices, information storage and processing devices (spintronics) [10]. To explore extensively the magnetic characteristics and using these properties in spintronic applications, the diluted magnetic semiconductors relied on III–V, II–VI, II–IV and IV–VI alloys which cations are carefully substituted by transition metals, were thoroughly the subjects of numerous theoretical and experimental investigations [11-12-13-14].

CaS is part of the group II-VI alkaline earth sulfide elements that crystallizes in rock salt structure, zinc blende, wurtzite and NiAs phases [15]. In general terms, the study of the electronic properties of materials makes it possible to investigate the electronic behavior of crystalline systems and to explain the nature of the chemical bonds between atoms. The electronic properties of CaS binary compound are broadly available in the literature, where calculations are performed by many researchers using the density functional theory (DFT) through various approximations. W.Y. Ching et al. [16] and Z. Charifi et al. [17] suggested from the first-principal calculations that the electronic band gap of CaS is about 3.2 eV and 2.39 eV, respectively. Y. Kaneko et al. [18-19] defined an indirect band gap located between Γ and X high-symmetry points for this binary compound around 4.434 eV, which is consistent with the theoretical investigations reported by M.S. Jin et al. [20] and J.G. Zhang et al. [21]. Moreover, Z.J. Chen et al. [22] proved from the band structure of CaS that the top of the valence band and the bottom of the conduction band are located at Γ point, suggesting a direct electronic band gap about 4.47 eV. However, Kravtsova et al. [23] and Saum et al. [24] stated that the electronic band gap width of CaS binary compound around 3.5 eV and 5.38 eV, respectively, without describing its nature. On the other hand, the half-metallic ferromagnetic properties of II–VI semiconductors based DMS compounds were extensively studied using the first-principle calculations, such us V and Cr doped CaS reported by Hamidane et al. [25-26], confirming the half-metallic ferromagnetism behavior in these compounds.

In the present research, we employed the first-principale theoretical methods based on spin-polarized density functional theory (SP-DFT), to study the thermodynamic stability and to investigate the structural parameters, the electronic properties, and the half metallic ferromagnetic features of $Ca_{1-x}Ti_xS$ ternary alloys.

### 2- Computational method

In this report, we have elaborated the characteristics of studied material by using the spin polarized density functional theory SP-DFT, which is operationalized by the full-potential linearized augmented plane wave method FP-LAPW [27] as implemented in Wien2K [28] ab-initio simulation code.

Perdew-Burk-Ernzerhof (PBE) parameterization of the generalized gradient approximation (GGA) approach is applied to describe the exchange-correlation (XC) potential and to estimate structural, electronic, and ferromagnetic properties of $Ca_{1-x}Ti_xS$ at various concentrations (x = 0.0625, 0.125, 0.25, 0.50 and 0.75). We also apply Tran Balaha modified Becke-Johnson exchange approach (TB − mBJ) [29] to exceed the problem posed by PBE-GGA [30] approximation which underestimates the value of the electronic band gap and also obtain optimal results.

The optimized CaS is stabilised in a rock-salt structure with space group Fm-3m (no. 225). Where the Ca and S atoms occupy respectively (0,0,0) and (0.5, 0.5, 0.5) positions. The supercell concept is employed to achieve $Ca_{1-x}Ti_xS$ diluted magnetic semiconductor, whose objective is to study the impact of transition metal (TM) atom substitution on Ca sites.

The product of the maximum modulus of reciprocal vector $K_{max}$ and the smallest of all atomic sphere radii $R_{MT}$ is expanded up to a constant value ($R_{MT}$ x $K_{max}$ = 8). Inside the muffin-tin spheres, the maximum partial wave is taken at $l_{max}$= 10. Whereas the extension of charge density is chosen at $G_{MAX}$ = 14 (a.u.)$^{-1}$. The muffin-tin radii (RMT) values for both Ca and Ti were 2.5 (a. u.), whereas for S is regarded as 2.2 (a. u.). In addition, an energy cutoff of −8.0 Ry is implemented to get the total energy eigenvalue convergence and avoid overlapping between the valence and core states. To ensure the correct results, a Monkhorst-Pack mesh [31-32] of 10 x10 x 10 k-points are implemented in the irreducible Brillouin zone for ferromagnetic and anti-ferromagnetic calculations. Self-consistency is supposed to be reached when the total energy difference of the system converged with an accuracy of $10^{-6}$ Ry, and the electron charge density was set at $10^{-4}$ e.

### 3- Results and discussion
#### 3-1- Structural properties

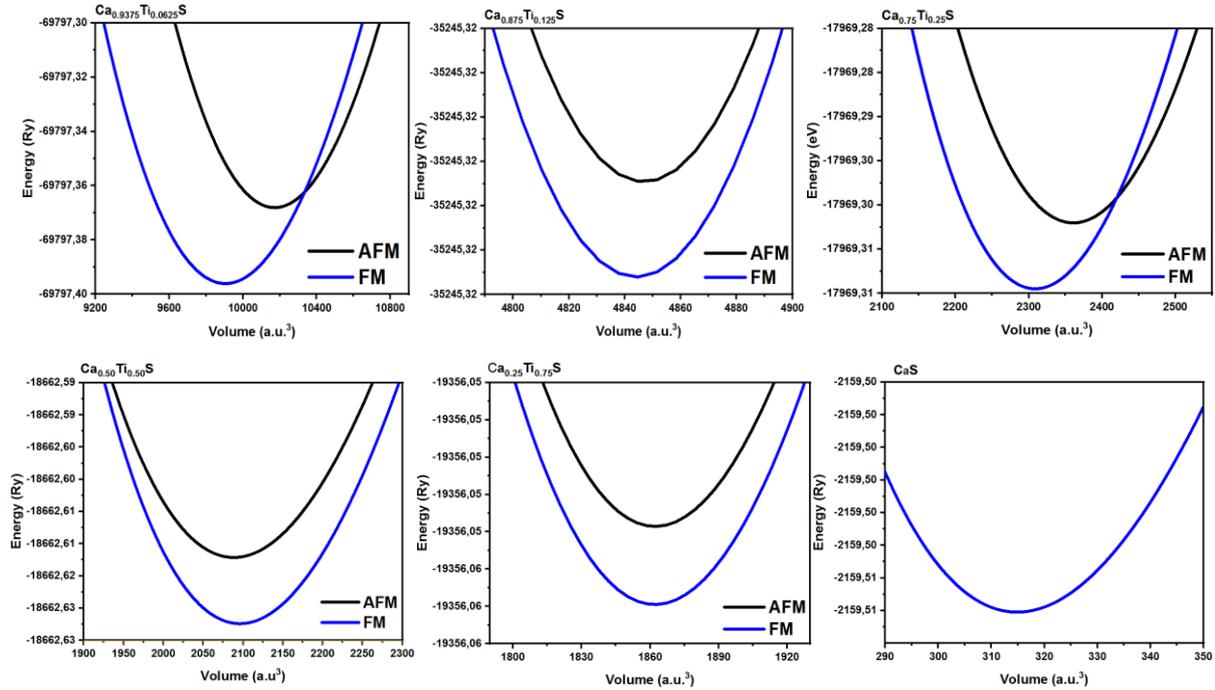

**Fig. 1.** Calculated total energy optimization as a function of volume per cell of the pristine CaS and the ternary systems $Ca_{1-x}Ti_xS$ (x = 0.0625, 0.125, 0.25, 0.50 and 0.75) in both ferromagnetic and anti-ferromagnetic states.

The calcium sulfide (CaS) is one of II-VI alkaline earth sulfide group, crystallizes in a rock-salt NaCl (B1) phase at ambient temperature [33-34] with space group of Fm3m (no.225) as mentioned above. In this structure, the atoms Ca and S are in the Wyckoff positions at 4a (0,0,0) and 4b (1/2, 1/2, 1/2), respectively. To evaluate the structural, electronic, and magnetic properties, we precisely performed the optimization of the total energy as a function of unit cell volume of CaS and $Ca_{1-x}Ti_xS$ compounds by the fitting of empirical Murnaghan's equation of state [35]. **Table 1**. Summarizes the expected structural parameters such as the equilibrium lattice constant (a₀), the bulk modulus (B), and its pressure derivative (B'), according to PBE-GGA approximation. We emphasize that for the pristine CaS, the results of our calculations are slightly lower than the available experimental data and agreed reasonably with other theoretical investigations as tabulated in **Table 1**. However, there are no realized experimental studies and theoretical works to be compared with structural parameters of $Ca_{1-x}Ti_xS$ alloys in the present work. It can further be noted from the table that substituting calcium by titanium atoms affects outstandingly the lattice constants (a₀) and produces a small distortion in the primitive unit cell. In fact, the observed reduction of lattice constants with increasing the concentrations x of the transition metal titanium (Ti), may be ascribed to the considerable difference between Ca and Ti ionic radius. As illustrated in **Fig. 1**, the calculated total energies for $Ca_{1-x}Ti_xS$ compounds in both ferromagnetic (FM) and anti-ferromagnetic (AFM) states, are plotted against the volume. A comparative analysis provides lower total energy of the ferromagnetic state than the total energy found in the anti-ferromagnetic state, indicating that the FM phase is more stable than the AFM phase. Furthermore, the FM nature for all compounds is verified by the positive values of energy differences between parallel and anti-parallel spin polarized states ($\Delta E = E_{AFM} - E_{FM}$), deduced from the self-consistent field (SCF) calculations (see **Table 2**). We conclude that even when the titanium concentration is gradually increased, the systems kept their ferromagnetic properties.

| Compounds | $a_0$(Å) | | | B(GPa) | | | B' | | |
|---|---|---|---|---|---|---|---|---|---|
| | Present | Exp. | Other | Present | Exp. | Other | Present | Exp. | Other |
| CaS | 5.7132 | 5.71 [49] | 5.68 [50] | 55.8933 | 64 [49] | 57.5 [50] | 3.8860 | 4.2 [49] | 3.8 [50] |
| $Ca_{0.9375}Ti_{0.0625}S$ | 5.7029 | | | 66.1334 | | | 1.7717 | | |
| $Ca_{0.875}Ti_{0.125}S$ | 5.6391 | | | 59.2631 | | | 4.1843 | | |
| $Ca_{0.75}Ti_{0.25}S$ | 5.5862 | | | 57.3976 | | | 1.9668 | | |
| $Ca_{0.50}Ti_{0.50}S$ | 5.3667 | | | 62.8531 | | | 4.0737 | | |
| $Ca_{0.25}Ti_{0.75}S$ | 5.1655 | | | 81.2757 | | | 5.0000 | | |

**Table 1.** Equilibrium lattice constant $a_0$(Å), bulk modulus B (GPa) and its pressure derivatives B' for CaS and $Ca_{1-x}Ti_xS$ compounds using PBE-GGA approximation.

Moreover, Curie temperature (Tc) can be computed using the classical model of Heisenberg in the framework of mean field approximation (MFA) [36], according to the subsequent relation:

$$T_c = \frac{2}{3K_\beta} \frac{\Delta E}{x} \qquad (1)$$

Where $\Delta E$ is the energy difference between FM and AFM states, $K_\beta$ is the Boltzman constant and x presents the doping concentration in the system. The predicted values of Curie temperature for the case of PBE-GGA approximation, and through modified Becke – Johnson exchange potential are featured in **Fig. 2**. It can be noted from this figure that the temperature increased progressively with the doping concentration of titanium (Ti), to reach the room temperature ferromagnetism at 12.5 % within TB – mBJ approach.

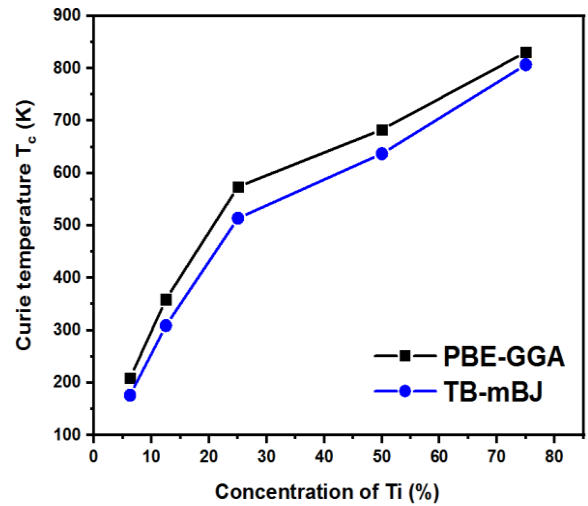

**Fig. 2.** Calculated Curie temperature (Tc) versus titanium concentration.

### 3-2- Defect formation energy

Density functional theory calculations can provide valuable estimates of the thermodynamic stability with appreciable accuracy. To investigate the stability of Ti-doped CaS compound, the formation energies are computed according to the following expression [37]:

$$E_{form} = E_T (Ca_{1-x}Ti_xS) - E_T(CaS) + n_{Ca}\mu_{Ca} - n_{Ti}\mu_{Ti} \qquad (2)$$

Where $E_T (Ca_{1-x}Ti_xS)$ represents the energy of doped system, $E_T(CaS)$ denotes the total energy of pure CaS supercell, $n_{Ti}$ and $n_{Ca}$ are the numbers of introduced and removed atom, respectively, while $\mu_{Ca}$ and $\mu_{Ti}$ exhibit the chemical potentials per atom of calcium and titanium bulk crystals. As depicted in **Fig. 3**, defect formation energy predicted from DFT calculations at PBE-GGA functional approximation decreased with respect to increasing concentrations of titanium. Whereas it seems to be more negative when the modified Becke-Johnson approach is implemented. However, the negative value of formation energy suggests that our compounds are thermodynamically stable in the ferromagnetic phase and can be readily prepared in the experiment.

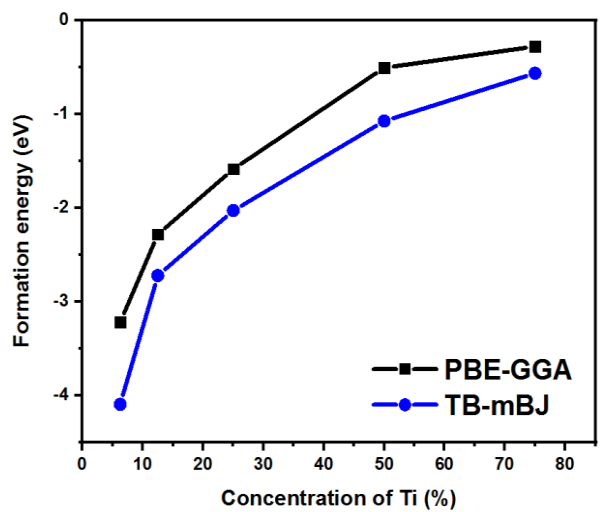

**Fig. 3.** Calculated formation energies as a function of Titanium concentration.

### 3-3- Band structure

The study of the electronic band structure and the density of state is essential for the semiconducting materials to identify its useful applications in various fields, by calculation of the energy band gap.

#### a- In the case of the pure material

By using PBE-GGA approximation, our calculations revealed that the pristine CaS is a semiconductor with a broad indirect band gap of 3.1522 eV, located between $\Gamma$ and X high-symmetry points, as displayed in **Fig. 6** and **Table 3**. To overcome the problem posed by PBE-GGA approximation which tends to underestimate the value of the electronic band gap [38-39], modified Becke-Johnson exchange potential approach is implemented. It is found that the electronic band gap is considerably enhanced and reached the value around 4.04545 eV. It is obvious from **Fig. 6** that the majority spin and the minority spin states are well overlapped suggesting the non-magnetic performance of the pristine CaS. It is worth mention that the electronic behavior of the pure compound is consistent with the theoretical calculations, and experimental measurement already reported (see **Table 3)**.

| Compounds | $E_g$ (eV) | $G_{HMF}^{\Gamma-\Gamma}$ (eV) | $G_{HM}$ (eV) | $\Delta E = E_{AFM} - E_{FM}$ (eV) | Behavior* |
|---|---|---|---|---|---|
| **PBE-GGA** | | | | | |
| $Ca_{0.9375}Ti_{0.0625}S$ | - | 2.2998 | 0.5336 | 0.001690 | HMF |
| $Ca_{0.875}Ti_{0.125}S$ | - | 2.2400 | 0.3962 | 0.005796 | HMF |
| $Ca_{0.75}Ti_{0.25}S$ | - | 2.0984 | 0.1011 | 0.018567 | HMF |
| $Ca_{0.50}Ti_{0.50}S$ | - | - | - | 0.044160 | Nearly HMF |
| $Ca_{0.25}Ti_{0.75}S$ | - | - | - | 0.080509 | M |
| **GGA + TB-mBJ** | | | | | |
| $Ca_{0.9375}Ti_{0.0625}S$ | - | 3.2144 | 1.1876 | 0.001420 | HMF |
| $Ca_{0.875}Ti_{0.125}S$ | - | 3.1608 | 1.0449 | 0.004992 | HMF |
| $Ca_{0.75}Ti_{0.25}S$ | $E_g^{R-\Gamma}(\uparrow) = 0.4155$; $E_g^{\Gamma-\Gamma}(\downarrow) = 2.8254$ | - | - | 0.016612 | MS |
| $Ca_{0.50}Ti_{0.50}S$ | $E_g^{R-X}(\uparrow) = 0.6281$; $E_g^{\Gamma-\Gamma}(\downarrow) = 2.8206$ | - | - | 0.041205 | MS |
| $Ca_{0.25}Ti_{0.75}S$ | - | 2.6759 | 0.1167 | 0.078234 | HMF |
| **GGA + U** | | | | | |
| $Ca_{0.9375}Ti_{0.0625}S$ | - | 3.0251 | 1.1446 | 0.001625 | HMF |
| $Ca_{0.875}Ti_{0.125}S$ | - | 2.8606 | 0.9660 | 0.004721 | HMF |
| $Ca_{0.75}Ti_{0.25}S$ | $E_g^{R-\Gamma}(\uparrow) = 0.4047$; $E_g^{\Gamma-\Gamma}(\downarrow) = 2.6825$ | - | - | 0.018510 | MS |
| $Ca_{0.50}Ti_{0.50}S$ | $E_g^{R-X}(\uparrow) = 0.5514$; $E_g^{\Gamma-\Gamma}(\downarrow) = 2.6750$ | - | - | 0.049407 | MS |
| $Ca_{0.25}Ti_{0.75}S$ | - | 2.4895 | 0.1598 | 0.088035 | HMF |

**Table 2.** Calculated band gap $E_g$ (eV), half-metallic gap $G_{HM}$ (eV) of minority spin states, half-metallic ferromagnetic gap $G_{HMF}^{\Gamma-\Gamma}$ (eV) and energy difference between ferromagnetic and anti-ferromagnetic states $\Delta E$ for $Ca_{1-x}Ti_xS$ compounds.

*The last column shows the type of materials: MS magnetic semiconductor, HMF half-metallic ferromagnetic, M metal in both spins.

| $E_g$ (eV) | | | | | |
|---|---|---|---|---|---|
| Present work | | Other theoretical works | | | Experimental |
| GGA-PBE | GGA + TB-mBJ | LDA | GGA-PBE | WC-GGA | |
| $E_g^{X-\Gamma} = 3.1522$ | $E_g^{X-\Gamma} = 4.04545$ | $E_g^{X-\Gamma} = 1.9$ [52] | $E_g^{X-\Gamma} = 2.1$ [52] | $E_g^{X-\Gamma} = 2.1$ [52] | $E_g^{X-\Gamma} = 5.8$ [51] |
| | | $E_g^{\Gamma-\Gamma} = 3.90$ [52] | $E_g^{\Gamma-\Gamma} = 4.25$ [52] | $E_g^{\Gamma-\Gamma} = 4.2$ [52] | $E_g^{\Gamma-\Gamma} = 5.34$ [51] |

**Table 3.** Calculated band gap $E_g$ (eV) for CaS compound using GGA-PBE and GGA + TB-mBJ.

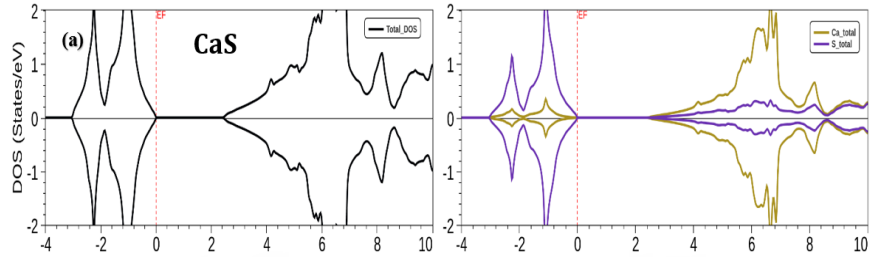
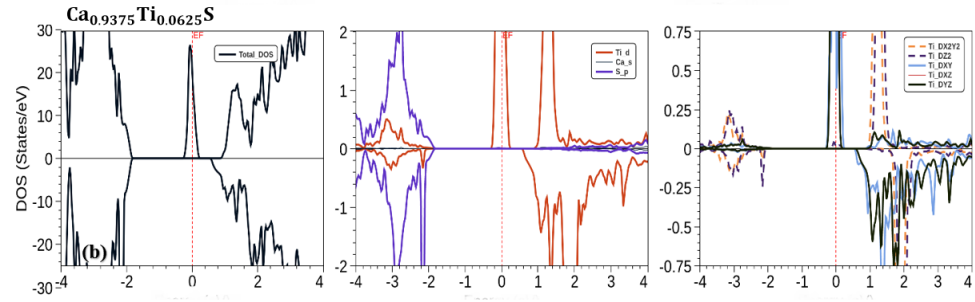
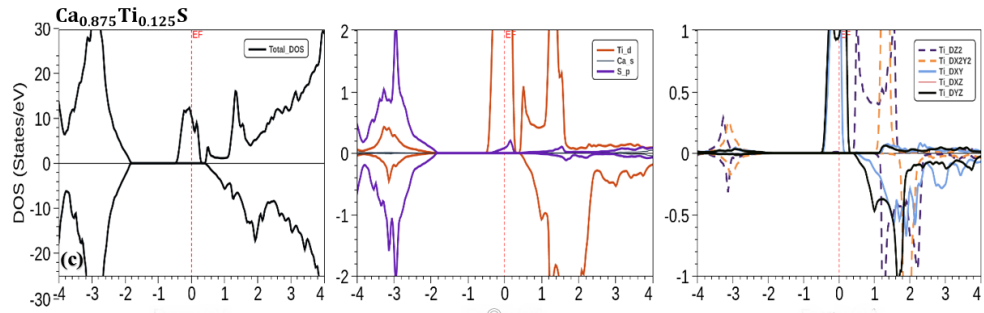
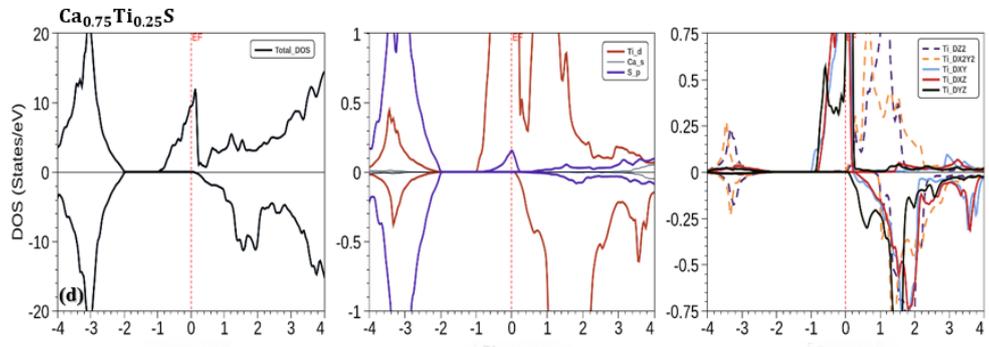
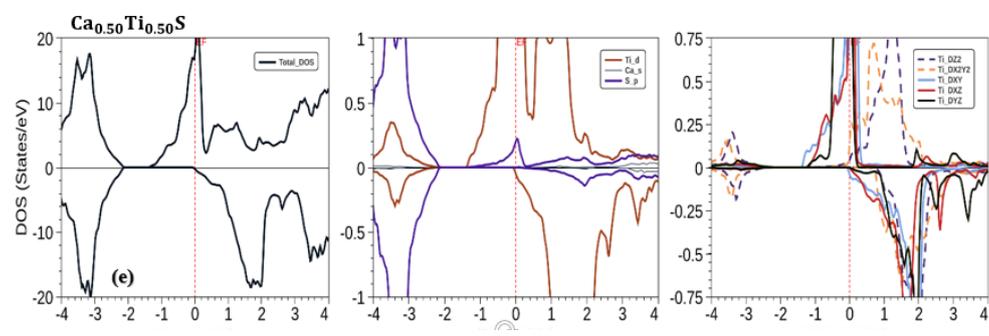

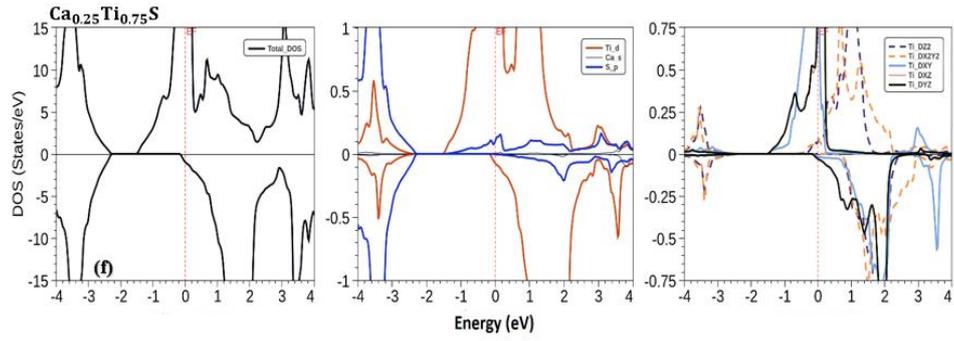

**Fig. 4.** Spin-polarized total and partial density of states of CaS and $Ca_{1-x}Ti_xS$ ternary alloys, using PBE-GGA approximation.

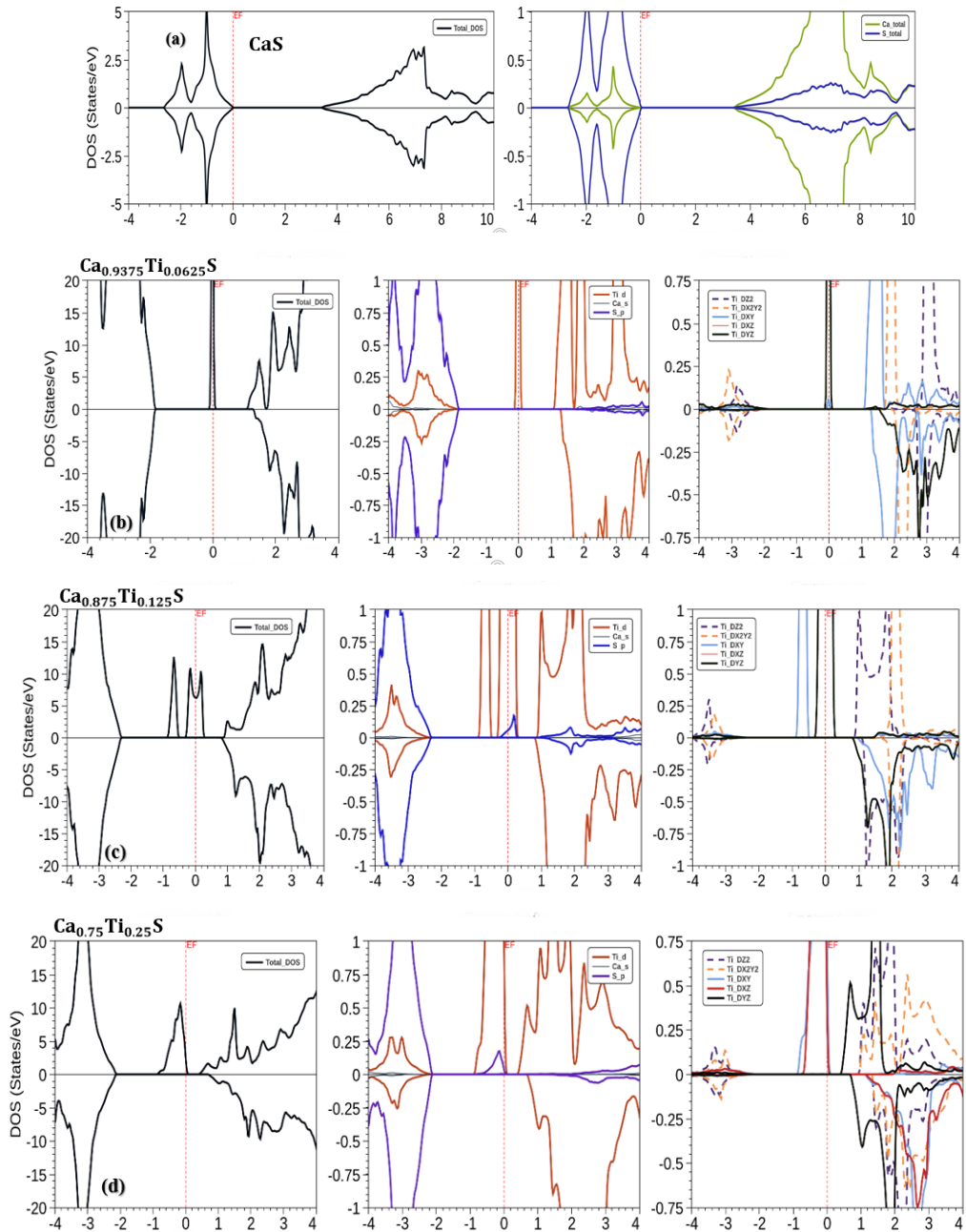

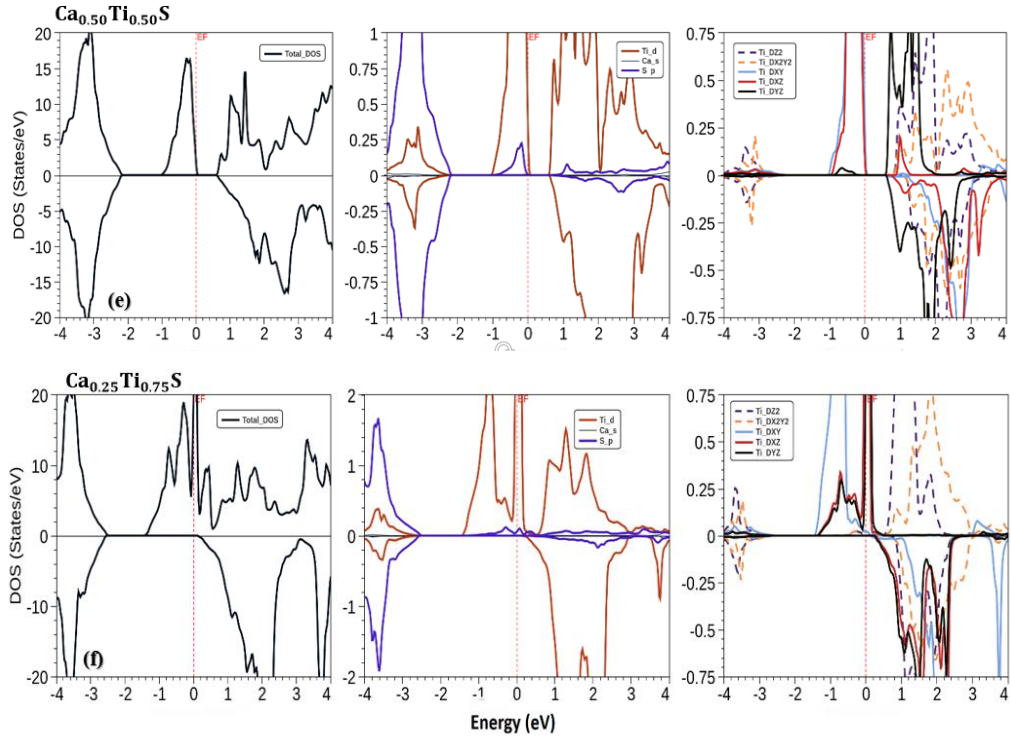

**Fig. 5.** Spin-polarized total and partial density of states of CaS and $Ca_{1-x}Ti_xS$ ternary alloys, using $TB-mBJ$ approach.

**b- In the case of the doped material**

The incorporation of titanium element (TM atom) in CaS compound improves the probability of producing new localized electronic states in the forbidden region, leads its nature to metallic and generates the $p-d$ hybridization between p and d orbitals of sulfur and titanium atoms respectively, resulting a ferromagnetic behavior in the system [40-41]. The spin-polarized band structures of ferromagnetic $Ca_{1-x}Ti_xS$ alloys (x = 0.0625, 0.125, 0.25, 0.50 and 0.75) are calculated at their equilibrium lattice constants by means of PBE-GGA and $TB-mBJ$ approximations. Along the high-symmetry directions in the first Brillouin zones, the spin-polarized band structures for $Ca_{0.9375}Ti_{0.0625}S$, $Ca_{0.875}Ti_{0.125}S$ and $Ca_{0.75}Ti_{0.25}S$ through the Generalized Gradient Approximation are displayed in **Fig. 6**. In fact, the minority spin (spin down channel) exhibits semiconducting demeanor, while the majority spin (spin up channel) presents a metallic behavior because the electronic states cross Fermi level ($E_f$). Therefore, the total contributions of both minority spin and majority spin states suggest the half metallic ferromagnetic (HMF) character with 100% spin-polarized for these compounds (see **Fig. 4**). Hence, the half metallic gaps ($G_{HM}$) and the half metallic ferromagnetic gaps ($G_{HMF}$) are created. Overall, the spin polarizations of the studied compounds are predicted in ferromagnetic state according to the following expression:

$$P(\%) = \frac{N_\uparrow(E_f) - N_\downarrow(E_f)}{N_\uparrow(E_f) + N_\downarrow(E_f)} \qquad (3)$$

Where $N_\uparrow(E_f)$ and $N_\downarrow(E_f)$ present the density of states for the spin up and the spin down channels, respectively, at Fermi level. The half metallic gap ($G_{HM}$) is defined as the minimum of the lowest energy conduction band and the absolute value of the highest energy valence band with respect to Fermi level of both spin states. While the half metallic ferromagnetic gap ($G_{HMF}$) corresponds to the energy difference between the upper fraction of the valence band and the minimum of the conduction band states around Fermi level at 0 eV [42-43-44] as clarified in **Fig. 6**. In the present case, the projected values of minority band gaps ($G_{HM}$) and ($G_{HMF}$) of the ternary alloys using PBE-GGA approximation, are tabulated in **Table 2**. It obviously appears from these results that for the spin down configuration, the expected values of the half metallic gaps ($G_{HM}$) decrease gradually with increasing concentration x of titanium due to the widening of 3d (Ti) states around the forbidden region ($E_f$). Whereas the top of the valence band and the bottom of conduction band are situated at $\Gamma$ high symmetry point providing a direct half metallic ferromagnetic gap ($G_{HMF}$). In contrast, the behavior of the $Ca_{0.50}Ti_{0.50}S$ and $Ca_{0.25}Ti_{0.75}S$ alloys is different from the others, which have a metallic nature in both spin states, as depicted in **Fig. 6**. The electronic behavior of materials is usually changed around Fermi level when the $TB - mBJ$ approach is implemented, as illustrated in **Fig. 6**. Further, for $Ca_{0.75}Ti_{0.25}S$ and $Ca_{0.50}Ti_{0.50}S$ ternary alloys, It is apparent that both the majority and minority spin states exhibit a semiconducting behavior, with an indirect band gap of about $E_g^{R-\Gamma} = 0.4155$ eV and $E_g^{R-X} = 0.6281$ eV, respectively, in the spin up channel, and a direct band gap around of $E_g^{\Gamma-\Gamma} = 2.8254$ eV and 2.8206 eV, respectively, in the spin down channel along $\Gamma-\Gamma$ direction (see **Table 2**). The spin-dependent half metallicity, corresponding to a conductor in the spin up states and a semiconductor in the spin down states is confirmed for $Ca_{0.9375}Ti_{0.0625}S$, $Ca_{0.875}Ti_{0.125}S$ and $Ca_{0.25}Ti_{0.75}S$ compounds, with 100% spin polarized behavior (presented in **Fig. 5**).

The half metallic gaps ($G_{HM}$) for these ternary alloys shown in **Table 2** correspond to the spin down configuration. In general terms, the broad values of $G_{HM}$ are a good sign of half metallic ferromagnets, as stated in reference [45]. Hence, the $Ca_{0.9375}Ti_{0.0625}S$ and $Ca_{0.875}Ti_{0.125}S$ with a greater half metallic gaps of about 1.1876 eV and 1.04494 eV, respectively (**Table 2**), are expected to be a stronger potential candidate for investigating the half metallic ferromagnetic properties, and they can be useful for spintronic device applications.

For further clarification, the main difference encountered in the band structures between PBE-GGA and the predictable approach technique named modified Becke – Johnson exchange potential (TB − mBJ), is the enhancement of the electronic band gaps values when the last approximation is implemented. In fact, Tran et al., have demonstrated that this form of approach develops over GGA and LDA potentials for the determination of band gaps values of the solids and the insulators [46]. Since bare DFT lacks accuracy on strongly correlated systems, a Hubbard corrected DFT energy functional is also used to treat such a strongly correlated system effectively. The Hubbard potential with different values of U is accurately optimized. At U = 6.5 eV, the findings suggest that there is no much discrepancy between GGA + TB − mBJ and GGA + U band structures (see **Table 2**). For each concentration of titanium, $Ca_{1-x}Ti_xS$ alloys kept the same behavior, which led to considering mBJ correction through the manuscript. Presumably, these approaches provide much improved results over the bare DFT for ground state properties such as energy band gaps, and provide a reasonable description of electronic properties. Our results may provide beneficial guidance for further theoretical research.

### 3-4- Density of state

To further analysis of substituting effect on the electronic band structures, and to explain the origin of ferromagnetism, we calculated the total density of states (TDOS) and the partial density of states (PDOS) near the Fermi levels within PBE-GGA and TB − mBJ approach, as illustrated in **Fig. 4** and **Fig. 5**. One can perceive that the plots confirm the achieved results in the previous section. For $Ca_{0.9375}Ti_{0.0625}S$, $Ca_{0.875}Ti_{0.125}S$ and $Ca_{0.75}Ti_{0.25}S$ at PBE-GGA potential, the minority spin states exhibit a semiconducting behavior with a broad electronic band gap, while a peak appears at Fermi level originating usually from Ti − 3d states hybridized slightly with S − p states, presenting a metallic behavior, which leads to the half metallic demeanor (**Fig. 4**). As described in the section above, $Ca_{0.9375}Ti_{0.0625}S$, $Ca_{0.875}Ti_{0.125}S$ and $Ca_{0.25}Ti_{0.75}S$ ternary alloys preserved the half metallic ferromagnets, when the modified Becke-Johnson exchange potential (TB − mBJ) is applied. In fact, the total density of states (TDOS) and the partial density of states (PDOS) show in that case, the existence of three regions:

The first region located in the valence band between -1 eV and - 4 eV occurred the mixed contributions of S − p and Ti − 3d states for both up and down spin channels. The p − d interaction between S − p and Ti − 3d orbitals in the valence band could create the magnetism in the studied compounds generating the ferromagnetism through the exchange splitting [53]. The second region at Fermi level, and the last region situated in the conduction band mainly arose from Ti − 3d orbitals (**Figs. 5**). The curves of the PDOS show that the Ti − 3d states split in to threefold degenerate high-lying states: $d_{xy}$, $d_{xz}$ and $d_{yz}$ (linear states), and twofold degenerate low-lying states: $d_{z2}$, $d_{x2}$-$d_{y2}$ (non-linear states) under the influence of the crystal field of neighboring sulfur atoms (**Fig. 4** and **Fig. 5**).

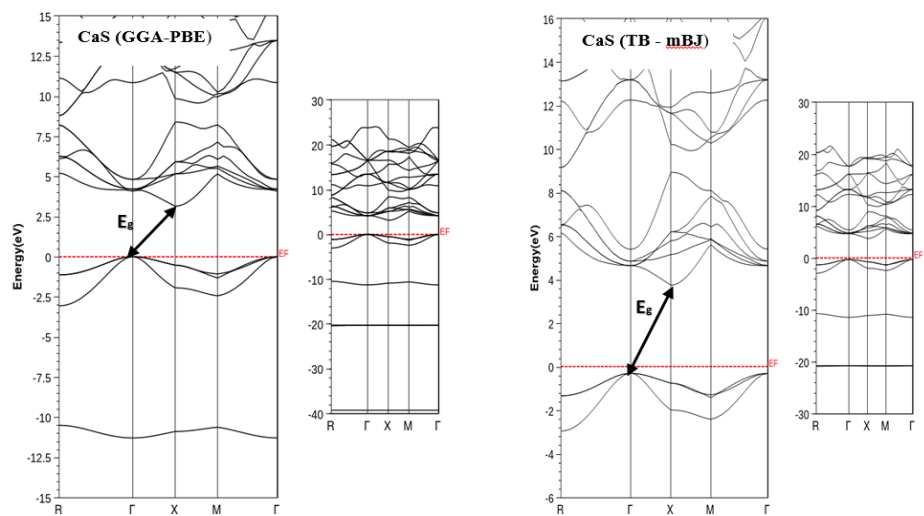

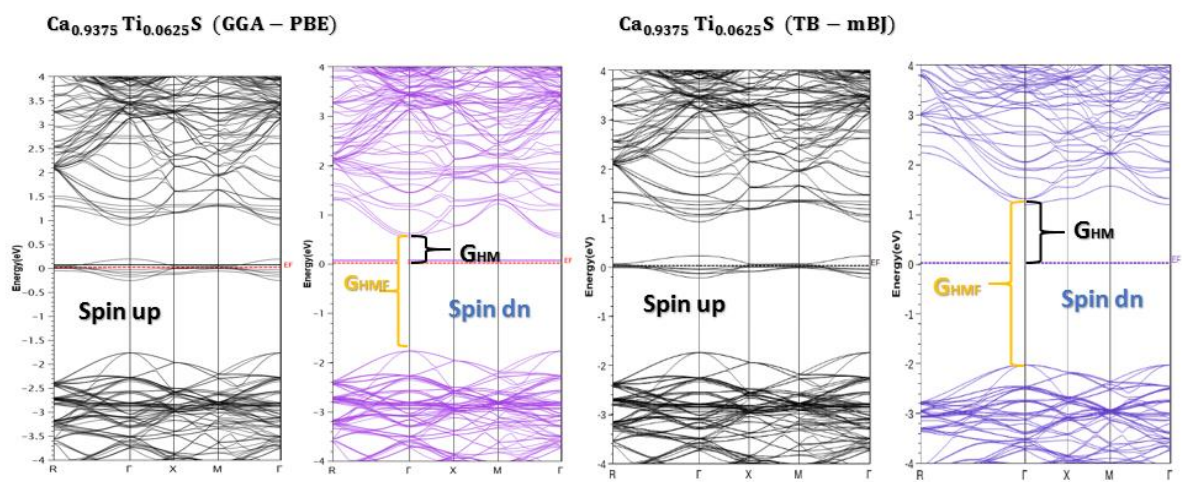

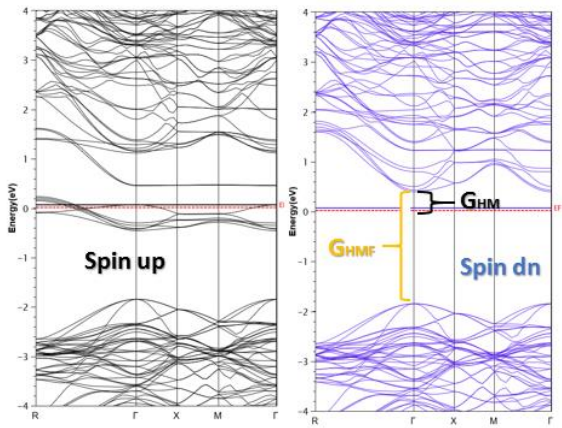
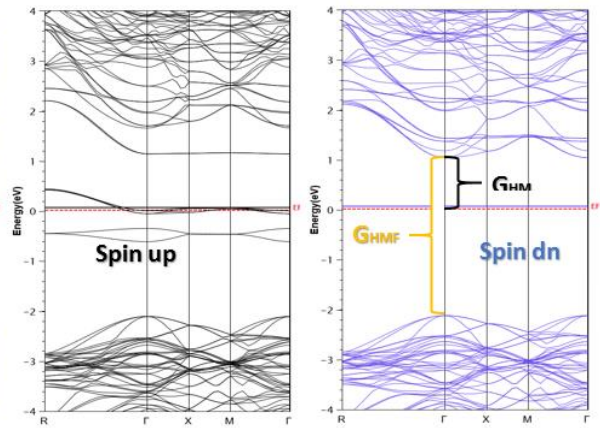
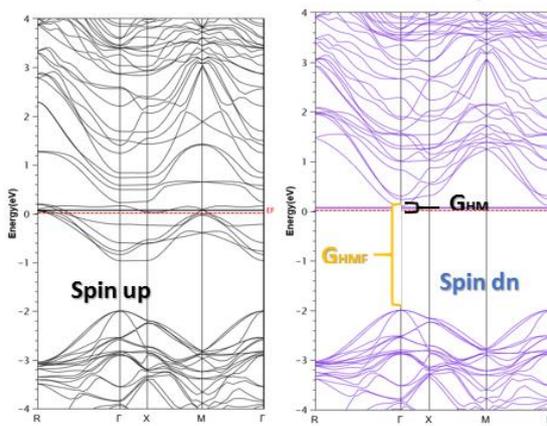
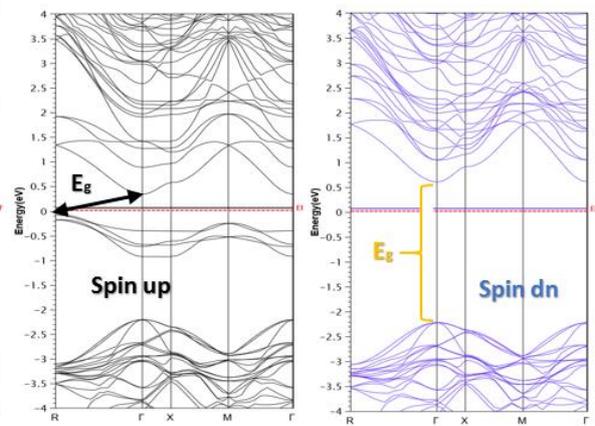
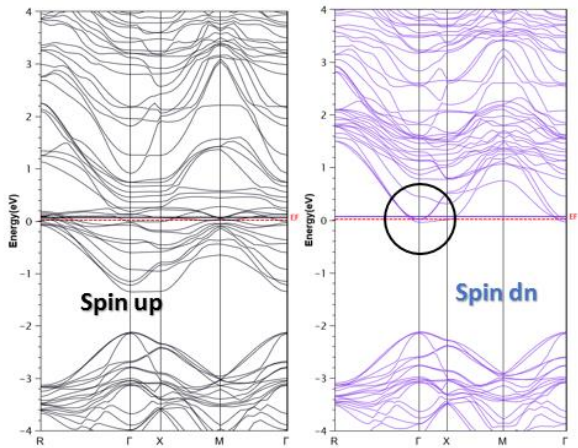
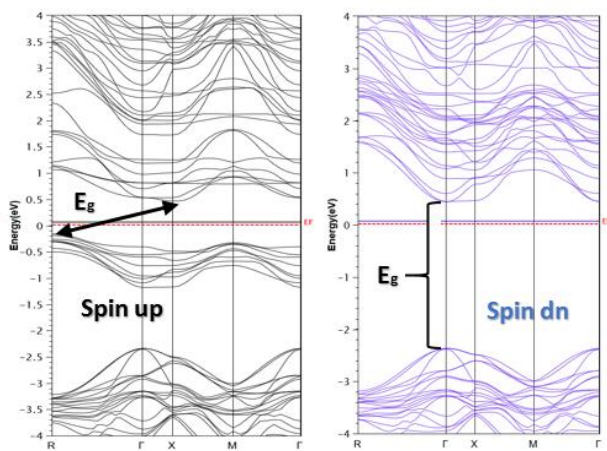

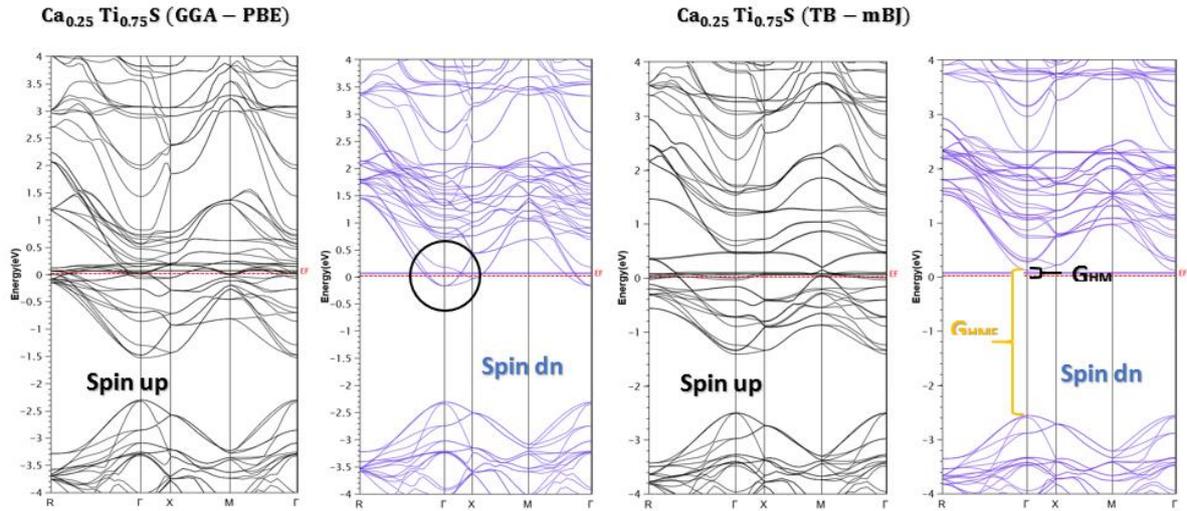

**Fig. 6.** Spin-polarized band structure of the pristine CaS and the equilibrium $Ca_{1-x}Ti_xS$ ternary alloys, using PBE-GGA and TB − mBJ approach.

### 3-5- Magnetic properties

The calculated total and local magnetic moments of $Ca_{1-x}Ti_xS$ (x = 0.0625, 0.125, 0.25, 0.50 and 0.75) ternary alloys by means of PBE-GGA and TB − mBJ approach are tabulated in **Table 4**. Obviously, the magnetization in $Ca_{1-x}Ti_xS$ compounds originates mainly from the Ti atom with outstanding contribution from the anion and interstitial sites. It is well known that the p − d hybridization leads to local magnetic moments at Ca, S and interstitial sites. The negative signs observed for the local magnetic moments of the sulfur anti-sites for all compounds, bear out that the interaction between S − p and Ti − 3d spins is anti-ferromagnetic. Whereas the ferromagnetic interaction can be justified by the same magnetic signs of Ti and Ca atoms (the positive magnetic moments). The computed magnetic moments for these ternary systems, became more substantial according to the gradual increase of titanium concentration.

Based on the mean field theory, we investigated using the band structures some important factors such as s − d exchange constants $N_{0\alpha}$, and p − d exchange constants $N_{0\beta}$. These significant parameters describe, respectively, the exchange interaction between the conduction electron carriers and Ti − 3d states (conduction band), and the exchange interaction between the holes and Ti − 3d states (valance band). The $N_{0\alpha}$, and $N_{0\beta}$ exchange constants can be computed using the following expressions [47]:

$$N_{0\alpha} = \frac{\Delta E_C}{x<S>} \quad (4)$$

$$N_{0\beta} = \frac{\Delta E_V}{x<S>} \quad (5)$$

Where $\Delta E_C = E(\downarrow) - E(\uparrow)$ and $\Delta E_V = E(\downarrow) - E(\uparrow)$ are the conduction band-edge and the valence band-edge spin-splittings, respectively, of $Ca_{1-x}Ti_xS$ compounds at the $\Gamma$ symmetry point of the band structures. $<S>$ represents the half total magnetic moment of Ti atom [47] and x is the concentration of titanium impurity. The calculated exchange constants for both potentials are tabulated in **Table 5**. One can clearly observe that the values of $N_{0\alpha}$, and $N_{0\beta}$ decrease as the concentration of the titanium atom increases, confirming the magnetic behavior of these ternary compounds, as mentioned in reference [48]. Additionally, the values of the exchange constants exhibit opposite signs. The positive signs of $N_{0\alpha}$ at all concentrations indicate the ferromagnetic exchange coupling between the conduction bands and the 3d states of titanium. While the negative values of $N_{0\beta}$ denote the anti-ferromagnetic exchange coupling between the valence bands and the $Ti - 3d$ states. The more negative $N_{0\beta}$ indicate a more effective spin down channel and this occurs in spin polarization systems [54].

| Compounds | $Ca_{0.9375}Ti_{0.0625}S$ | $Ca_{0.875}Ti_{0.125}S$ | $Ca_{0.75}Ti_{0.25}S$ | $Ca_{0.50}Ti_{0.50}S$ | $Ca_{0.25}Ti_{0.75}S$ |
|---|---|---|---|---|---|
| **PBE-GGA** | | | | | |
| $M_{tot}$ | 3.78878 | 3.98710 | 4.03081 | 7.99573 | 11.99845 |
| $M_{Ti}$ | 3.09126 | 3.09942 | 3.16841 | 6.42924 | 9.75980 |
| $M_{Ca}$ | 0.04632 | 0.08598 | 0.05558 | 0.06107 | 0.04531 |
| $M_S$ | -0.11984 | -0.10089 | -0.04194 | -0.09359 | -0.14725 |
| Interstitial | 0.77104 | 0.90260 | 0.84876 | 1.59901 | 2.34059 |
| **GGA + TB-mBJ** | | | | | |
| $M_{tot}$ | 4.00 | 4.00 | 4.00 | 8.00 | 11.99 |
| $M_{Ti}$ | 3.29293 | 3.26162 | 3.37455 | 6.75916 | 9.99234 |
| $M_{Ca}$ | 0.05892 | 0.05514 | 0.02682 | 0.03691 | 0.03544 |
| $M_S$ | -0.02174 | -0.03609 | 0.03742 | 0.06629 | 0.02801 |
| Interstitial | 0.67656 | 0.71935 | 0.56123 | 1.13764 | 1.94397 |

**Table 4.** Calculated results of total magnetic moment and local magnetic moment for $Ca_{1-x}Ti_xS$ compounds.

| Compounds | $E_v(\uparrow)$ | $E_v(\downarrow)$ | $E_c(\uparrow)$ | $E_c(\downarrow)$ | $\Delta E_v$ | $\Delta E_c$ | $N_{0\alpha}$ | $N_{0\beta}$ |
|---|---|---|---|---|---|---|---|---|
| **PBE-GGA** | | | | | | | | |
| $Ca_{0.9375}Ti_{0.0625}S$ | 0.00 | -1.771 | 0.00 | 0.5336 | -1.771 | 0.533 | 5.525 | -18.34 |
| $Ca_{0.875}Ti_{0.125}S$ | 0.00 | -1.843 | 0.00 | 0.39626 | -1.843 | 0.396 | 2.051 | -9.547 |
| $Ca_{0.75}Ti_{0.25}S$ | 0.00 | -1.992 | 0.00 | 0.10113 | -1.992 | 0.101 | 0.256 | -5.044 |
| $Ca_{0.50}Ti_{0.50}S$ | 0.00 | 0.00 | 0.00 | 0.00 | 0.00 | 0.00 | - | - |
| $Ca_{0.25}Ti_{0.75}S$ | 0.00 | 0.00 | 0.00 | 0.00 | 0.00 | 0.00 | - | - |
| **GGA + TB-mBJ** | | | | | | | | |
| $Ca_{0.9375}Ti_{0.0625}S$ | 0.00 | -2.026 | 0.00 | 1.184 | -2.026 | 1.187 | 9.500 | -16.21 |
| $Ca_{0.875}Ti_{0.125}S$ | 0.00 | -2.115 | 0.00 | 1.044 | -2.115 | 1.044 | 4.179 | -8.463 |
| $Ca_{0.75}Ti_{0.25}S$ | -0.071 | -2.216 | 0.343 | 0.608 | -2.145 | 0.265 | 0.529 | -4.290 |
| $Ca_{0.50}Ti_{0.50}S$ | -0.157 | -2.361 | 0.470 | 0.442 | -2.203 | -0.027 | -0.014 | -1.101 |
| $Ca_{0.25}Ti_{0.75}S$ | 0.00 | -2.558 | 0.00 | 0.116 | -2.558 | 0.117 | 0.0045 | -0.426 |

**Table 5.** Calculated conduction and valence band-edge spin splitting $\Delta E_c$, $\Delta E_v$ and exchange constants $N_{0\alpha}$, $N_{0\beta}$ of $Ca_{1-x}Ti_xS$ compounds for all concentrations.

# 4- Conclusion

In this paper, we investigated carefully by the mean of the spin-polarized density functional theory, the structural, electronic, and magnetic properties of $Ca_{1-x}Ti_xS$ ternary alloys. The most important findings of the present work are as follows:

- The equilibrium lattice parameter decreases gradually with increasing concentration of titanium impurity due to the difference between the Ca and Ti atomic radii.
- The stability of the ferromagnetic states is more than the anti-ferromagnetic states, while the negative values of the formation energies irrespective of the substitution concentration lead to a thermodynamic stability.
- The electronic properties calculated using PBE-GGA approximation confirm a half metallic ferromagnetic (HMF) behavior for $Ca_{0.9375}Ti_{0.0625}S$, $Ca_{0.875}Ti_{0.125}S$ and $Ca_{0.75}Ti_{0.25}S$ compounds. However, $Ca_{0.50}Ti_{0.50}S$ and $Ca_{0.25}Ti_{0.75}S$ alloys exhibit a metallic nature.
- When $TB-mBJ$ is implemented, $Ca_{0.75}Ti_{0.25}S$ and $Ca_{0.50}Ti_{0.50}S$ ternary compounds maintain a semiconducting nature, while $Ca_{0.9375}Ti_{0.0625}S$, $Ca_{0.875}Ti_{0.125}S$ and $Ca_{0.25}Ti_{0.75}S$ exhibit a half metallic demeanor with 100 % spin polarization at Fermi level.
- The total magnetic moment rises with the increase of titanium concentration, which is mainly provided by Ti atoms.
- $Ca_{1-x}Ti_xS$ ternary alloys at low concentrations (x = 0.0625 and 0.125) show a higher half-metallic gap ($G_{HM}$). Hence, these compounds are predicted to be potential materials for spintronic device applications.

Our theoretical results need to be confirmed by the experiments and further theoretical investigations.

**Acknowledgments**

The authors are thankful to Prof. P. Blaha and Prof. K. Schwarz at Wien Technical University for the Wien2k package and the group of WIEN2K for useful discussions.